\documentclass[a4paper]{jpconf}
\usepackage{graphicx}
\usepackage{bm}
\begin{document}
\title{Heavy quarkonia: the beauty and the beasts}
\author{George Rupp$^1$ and Eef van Beveren$^2$}
\address{$^1$ Centro de F\'{\i}sica e Engenharia de Materiais
Avan\c{c}ados, Instituto Superior T\'{e}cnico, Universidade de Lisboa,
P-1049-001, Portugal}
\address{$^2$ Centro de F\'{\i}sica da UC, Departamento de
F\'{\i}sica, Universidade de Coimbra, P-3004-516, Portugal}
\ead{george@ist.utl.pt eef@uc.pt}
\begin{abstract}
New enhancements in the charmonium and bottomonium spectra observed since
2003 are very briefly reviewed. Special attention is paid to
$\chi_{c1}(3872)$ (formerly $X(3872)$) owing to its remarkable proximity
to the $\bar{D}^{\star0}\!D^0$ threshold, which allows 
modelling as a quasibound axial-vector $c\bar{c}$ state with a large
$\bar{D}^{\star0}\!D^0$ admixture. In contrast, the interpretation
of many other charmonium-like and bottomonium-like states is still very
controversial and some may not even correspond to genuine resonances.
Accordingly, several entries in the PDG tables have been  wildly changing
over the years. Three representative states are reviewed here as
non-resonant enhancements due to threshold effects, viz.\ $\psi(4260)$,
$\psi(4660)$, and $\Upsilon(10580)$.
\end{abstract}
\section{Introduction}
Meson spectroscopy has witnessed a true revolution since the year
2003, when a first unusual charmonium-like state was observed
\cite{PRL91p262001} by the Belle Collaboration, extremely close to the
$D^0\bar{D}^{\star0}$ threshold at 3872~MeV. For this reason, the
collaboration suggested the state might correspond to a (quote) \em loosely
bound $D\bar{D}^\star$ multiquark ``molecular'' state, as proposed by some
authors \em \/(unquote). The designation $X(3872)$ employed in
Ref.~\cite{PRL91p262001} was then also adopted by the PDG in 2004
\cite{PDG04} and maintained till 2017. In the very recent 2018
edition \cite{PDG18} the name was finally changed into $\chi_{c1}(3872)$,
based on unmistakable evidence favouring $J^{PC}=1^{++}$ quantum numbers.
However, despite being the obvious candidate for $\chi_{c1}(2P)$, its
spectroscopic assignment is still controversial, with several authors
insisting on a non-$c\bar{c}$ configuration, the molecular hypothesis
being the most popular. Whatever the description, though, $\chi_{c1}(3872)$
is absolutely unique among the newly discovered heavy quarkonia, owing to
its accurately known mass, very small width, and well-established $J^{PC}$.
This allows a detailed and selective modelling, as well as dedicated
experiments aiming at pinning down its properties even more precisely.
In Sec.~3 below we shall succinctly revisit some of our model results on
$\chi_{c1}(3872)$.

On the other hand, many of the other enigmatic heavy-quarkonium states
observed in recent years are much less firmly established and some are
even questionable as genuine resonances. Checking their entries in the PDG
tables since 2004, one observes frequent changes of listed masses and
names, with an occasional state being eliminated altogether (see Sec.~2
below for details). Most of the model activity has focused on
charged states like e.g.\ $X(3900)^\pm$ (now \cite{PDG18} called
$Z_c(3900)^\pm$) and $X(10610)^\pm$ (now \cite{PDG18} 
$Z_b(10610)^\pm$), which cannot be purely $c\bar{c}$ or $b\bar{b}$, of
course. For a short review on such ``$XYZ$'' states, see 
Ref.~\cite{JPSCP17p111004}, in particular concerning non-resonant
assignments. Another very puzzling state is $X(4260)$ (now \cite{PDG18}
$\psi(4260)$) which has not been observed in \em any \em \/open-charm
decay channel. In Sec.~4 we shall very briefly review our non-resonant
model description of $\psi(4260)$), as well as that of $\psi(4660)$ and
$\Upsilon(10580)$.
\section{Charmonium and bottomonium states in the PDG tables since 2002}
Up to the 2002 PDG edition \cite{PDG02}, the list of charmonia and
bottomonia had remained largely unaltered, with all states being considered
compatible with mainstream quark models. Things changed dramatically in
2003 with the discovery \cite{PRL91p262001} of $X(3872)$, whose mass is
almost 100 MeV lower than was expected then (also see Sec.~3 below). Over
the following years many more new charmonium-like states were observed and
included in the PDG listings, as summarised next.

In the following charmonium and bottomonium enumerations, normal entries
refer to states already present in the preceding PDG edition, boldface
ones to new entries, designations in quotation marks to mere name changes,
and states in square brackets to freshly removed entries. \\[2mm]
\underline{\bf Charmonium} \\[2mm]
{\bf2002:}\hfill\parbox[t]{14.6cm}{
$\eta_c(1S)$ $J/\psi(1S)$ $\chi_{c0}(1P)$ $\chi_{c1}(1P)$ $h_c(1P)$
$\chi_{c2}(1P)$ $\eta_c(2S)$ $\psi(2S)$ $\psi(3770)$ $\psi(3836)$
$\psi(4040)$ $\psi(4160)$ $\psi(4415)$} \\[1.5mm]
{\bf2004:}\hfill\parbox[t]{14.6cm}{
$\eta_c(1S)$ $J/\psi(1S)$ $\chi_{c0}(1P)$ $\chi_{c1}(1P)$ $h_c(1P)$
$\chi_{c2}(1P)$ $\eta_c(2S)$ $\psi(2S)$ $\psi(3770)$ $\psi(3836)$
$\bm{X(3872)}$ $\psi(4040)$ $\psi(4160)$ $\psi(4415)$} \\[1.5mm]
{\bf2006:}\hfill\parbox[t]{14.6cm}{
$\eta_c(1S)$ $J/\psi(1S)$ $\chi_{c0}(1P)$ $\chi_{c1}(1P)$ $h_c(1P)$
$\chi_{c2}(1P)$ $\eta_c(2S)$ $\psi(2S)$ $\psi(3770)$
$\left[\psi(3836)\right]$
$X(3872)$ $\bm{\chi_{c2}(2P)}$ $\bm{X(3940)}$ $\psi(4040)$ $\psi(4160)$
$\bm{Y(4260)}$ $\psi(4415)$} \\[1.5mm]
{\bf2008:}\hfill\parbox[t]{14.8cm}{
$\eta_c(1S)$ $J/\psi(1S)$ $\chi_{c0}(1P)$ $\chi_{c1}(1P)$ $h_c(1P)$
$\chi_{c2}(1P)$ $\eta_c(2S)$ $\psi(2S)$ $\psi(3770)$
$X(3872)$ $\chi_{c2}(2P)$ $X(3940)$ $\bm{X(3945)}$ $\psi(4040)$ $\psi(4160)$
``$X(4260)$'' $\bm{X(4360)}$ $\psi(4415)$} \\[1.5mm]
{\bf2010:}\hfill\parbox[t]{14.8cm}{
$\eta_c(1S)$ $J/\psi(1S)$ $\chi_{c0}(1P)$ $\chi_{c1}(1P)$ $h_c(1P)$
$\chi_{c2}(1P)$ $\eta_c(2S)$ $\psi(2S)$ $\psi(3770)$
$X(3872)$ $\chi_{c2}(2P)$ $X(3940)$ $X(3945)$ $\psi(4040)$
$\bm{X(4050)^{\pm}}$ $\bm{X(4140)}$  $\psi(4160)$ $\bm{X(4160)}$
$\bm{X(4250)^{\pm}}$ $X(4260)$ $\bm{X(4350)}$ $X(4360)$
$\psi(4415)$ $\bm{X(4430)^{\pm}}$ $\bm{X(4660)}$} \\[1.5mm]
{\bf2012:}\hfill\parbox[t]{14.8cm}{
$\eta_c(1S)$ $J/\psi(1S)$ $\chi_{c0}(1P)$ $\chi_{c1}(1P)$ $h_c(1P)$
$\chi_{c2}(1P)$ $\eta_c(2S)$ $\psi(2S)$ $\psi(3770)$ $X(3872)$
$\bm{X(3915)}$ $\chi_{c2}(2P)$ $X(3940)$
$\left[X(3945)\right]$
$\psi(4040)$ $X(4050)^{\pm}$ $X(4140)$  $\psi(4160)$ $X(4160)$
$X(4250)^{\pm}$ $X(4260)$ $X(4350)$ $X(4360)$
$\psi(4415)$ $X(4430)^{\pm}$ $X(4660)$} \\[1.5mm]
{\bf2014:}\hfill\parbox[t]{14.8cm}{
$\eta_c(1S)$ $J/\psi(1S)$ $\chi_{c0}(1P)$ $\chi_{c1}(1P)$ $h_c(1P)$
$\chi_{c2}(1P)$ $\eta_c(2S)$ $\psi(2S)$ $\psi(3770)$ $\bm{X(3823)}$
$X(3872)$ $\bm{X(3900)^{\pm}}$ $\bm{X(3900)^0}$ $X(3915)$
$\bm{\chi_{c0}(2P)}$ $\chi_{c2}(2P)$ $X(3940)$ $\bm{X(4020)^{\pm}}$
$\psi(4040)$ $X(4050)^{\pm}$ $X(4140)$  $\psi(4160)$ $X(4160)$
$X(4250)^{\pm}$ $X(4260)$ $X(4350)$ $X(4360)$
$\psi(4415)$ $X(4430)^{\pm}$ $X(4660)$} \\[1.5mm]
{\bf2016:}\hfill\parbox[t]{14.8cm}{
$\eta_c(1S)$ $J/\psi(1S)$ $\chi_{c0}(1P)$ $\chi_{c1}(1P)$ $h_c(1P)$
$\chi_{c2}(1P)$ $\eta_c(2S)$ $\psi(2S)$ $\psi(3770)$ ``$\psi(3823)$''
$X(3872)$ $X(3900)^{\pm}$
$\left[X(3900)^0\right]$
$X(3915)$
$\left[\chi_{c0}(2P)\right]$
$\chi_{c2}(2P)$ $X(3940)$ $X(4020)^{\pm}$ $\psi(4040)$ $X(4050)^{\pm}$
$\bm{X(4055)^{\pm}}$ $X(4140)$ $\psi(4160)$ $X(4160)$ $\bm{X(4200)^{\pm}}$
$\bm{X(4230)}$ $\bm{X(4240)^{\pm}}$ $X(4250)^{\pm}$ $X(4260)$ $X(4350)$
$X(4360)$ $\psi(4415)$ $X(4430)^{\pm}$ $X(4660)$} \\[1.5mm]
{\bf2018:}\hfill\parbox[t]{14.8cm}{
$\eta_c(1S)$ $J/\psi(1S)$ $\chi_{c0}(1P)$ $\chi_{c1}(1P)$ $h_c(1P)$
$\chi_{c2}(1P)$ $\eta_c(2S)$ $\psi(2S)$ $\psi(3770)$ ``$\psi_2(3823)$''
$\bm{\chi_{c0}(3860)}$ ``$\chi_{c1}(3872)$'' ``$Z_c(3900)$'' $X(3915)$
``$\chi_{c2}(3930)$'' $X(3940)$ $X(4020)^{\pm}$ $\psi(4040)$ $X(4050)^{\pm}$
$X(4055)^{\pm}$ ``$\chi_{c1}(4140)$'' $\psi(4160)$ $X(4160)$ ``$Z_c(4200)$''
``$\psi(4230)$'' ``$R_{c0}(4240)$'' $X(4250)^{\pm}$ ``$\psi(4260)$''
$\bm{\chi_{c1}(4274)}$ $X(4350)$ ``$\psi(4360)$'' $\bm{\psi(4390)}$ 
$\psi(4415)$ ``$Z_c(4430)$'' $\bm{\chi_{c0}(4500)}$ ``$\psi(4660)$''
$\bm{\chi_{c0}(4700)}$} \\[2mm]
The above PDG entries over the period 2002--2018 show an increasingly
confusing situation of well-established charmonia, puzzling charmonium-like
resonances though with non-exotic quantum numbers, and clear
non-$c\bar{c}$-only states, either for being charged or having exotic
quantum numbers (cf.~$R_{c0}(4240)$, with tentative $J^{PC}=0^{--}$
\cite{PDG18}). Among the several questionable assignments made by the
PDG \cite{PDG18}, one should stress the regressive designation of $X(3915)$,
previously named $\chi_{c0}(2P)$ \cite{PDG14} or $\chi_{c0}(3915)$
\cite{PDG15}, and to be contrasted with the new $\chi_{c0}(4500)$ and
$\chi_{c0}(4700)$. \\
\underline{\bf Bottomonium} \\[2mm]
{\bf2002:}\hfill\parbox[t]{14.8cm}{
$\eta_b(1S)$ $\Upsilon(1S)$ $\chi_{b0}(1P)$ $\chi_{b1}(1P)$ $\chi_{b2}(1P)$
$\Upsilon(2S)$ $\chi_{b0}(2P)$ $\chi_{b1}(2P)$ $\chi_{b2}(2P)$
$\Upsilon(3S)$ $\Upsilon(4S)$ $\Upsilon(10860)$ $\Upsilon(11020)$} \\[1.5mm]
{\bf2006:}\hfill\parbox[t]{14.8cm}{
$\eta_b(1S)$ $\Upsilon(1S)$ $\chi_{b0}(1P)$ $\chi_{b1}(1P)$ $\chi_{b2}(1P)$
$\Upsilon(2S)$ $\bm{\Upsilon(1D)}$ $\chi_{b0}(2P)$ $\chi_{b1}(2P)$
$\chi_{b2}(2P)$ $\Upsilon(3S)$ $\Upsilon(4S)$ $\Upsilon(10860)$
$\Upsilon(11020)$} \\[1.5mm]
{\bf2012:}\hfill\parbox[t]{14.8cm}{
$\eta_b(1S)$ $\Upsilon(1S)$ $\chi_{b0}(1P)$ $\chi_{b1}(1P)$ $\bm{h_b(1P)}$
$\chi_{b2}(1P)$ $\Upsilon(2S)$ $\Upsilon(1D)$ $\chi_{b0}(2P)$
$\chi_{b1}(2P)$ $\bm{h_b(2P)}$ $\chi_{b2}(2P)$ $\Upsilon(3S)$
$\bm{\chi_b(3P)}$ $\Upsilon(4S)$ $\bm{X(10610)^{\pm}}$ 
$\bm{X(10650)^{\pm}}$ $\Upsilon(10860)$ $\Upsilon(11020)$} \\[1.5mm]
{\bf2014:}\hfill\parbox[t]{14.8cm}{
$\eta_b(1S)$ $\Upsilon(1S)$ $\chi_{b0}(1P)$ $\chi_{b1}(1P)$ $h_b(1P)$
$\chi_{b2}(1P)$ $\bm{\eta_b(2S)}$ $\Upsilon(2S)$ $\Upsilon(1D)$
$\chi_{b0}(2P)$ $\chi_{b1}(2P)$ $h_b(2P)$ $\chi_{b2}(2P)$
$\Upsilon(3S)$ $\chi_b(3P)$ $\Upsilon(4S)$ $X(10610)^{\pm}$ 
$\bm{X(10610)^0}$ $X(10650)^{\pm}$ $\Upsilon(10860)$ $\Upsilon(11020)$}
\\[1.5mm]
{\bf2016:}\hfill\parbox[t]{14.8cm}{
$\eta_b(1S)$ $\Upsilon(1S)$ $\chi_{b0}(1P)$ $\chi_{b1}(1P)$ $h_b(1P)$
$\chi_{b2}(1P)$ $\eta_b(2S)$ $\Upsilon(2S)$ $\Upsilon(1D)$
$\chi_{b0}(2P)$ $\chi_{b1}(2P)$ $h_b(2P)$ $\chi_{b2}(2P)$
$\Upsilon(3S)$ ``$\chi_{b1}(3P)$'' $\Upsilon(4S)$ $X(10610)^{\pm}$ 
$X(10610)^0$ $X(10650)^{\pm}$ $\Upsilon(10860)$ $\Upsilon(11020)$}
\\[1.5mm]
{\bf2018:}\hfill\parbox[t]{14.8cm}{
$\eta_b(1S)$ $\Upsilon(1S)$ $\chi_{b0}(1P)$ $\chi_{b1}(1P)$ $h_b(1P)$
$\chi_{b2}(1P)$ $\eta_b(2S)$ $\Upsilon(2S)$ ``$\Upsilon_2(1D)$''
$\chi_{b0}(2P)$ $\chi_{b1}(2P)$ $h_b(2P)$ $\chi_{b2}(2P)$
$\Upsilon(3S)$ $\chi_{b1}(3P)$ $\Upsilon(4S)$ ``$Z_b(10610)$'' 
``$Z_b(10650)$'' $\Upsilon(10860)$ $\Upsilon(11020)$}
\\[2mm]
In the $b\bar{b}$ sector the successive PDG updates have followed a more 
conservative pattern, also owing to the reduced number of newly discovered
states. (Only PDG editions with changes given here). Nevertheless, the
designation $\chi_b(3P)$ \cite{PDG12,PDG14} is
incomprehensible and even unacceptable, as it refers to 3 different
bottomonia, with $J=0, 1,$ or 2. Impossibility to disentangle them, due
to insufficient experimental resolution \cite{ARXIV12041984}, should have
imposed an ``$X$'' naming.
\section{$\bm{\chi_{c1}(3872)}$ as a unitarised $\bm{\chi_{c1}(2P)}$ state}
Hardly any other meson has been attracting more attention from model
builders than $\chi_{c1}(3872)$ (alias $X(3872)$, as it was called
\cite{PDG16} before the 2018 PDG edition \cite{PDG18}. Among the various
interpretations of this meson, the molecular picture is probably the most
popular. However, any non-exotic hypothetical molecular state will inevitably
mix with a bare $c\bar{c}$ state having the same quantum numbers, via light
$q\bar{q}$ creation/annihilation and the $^{3\!}P_0$ mechanism, even if the
$c\bar{c}$ state lies at a quite different energy \cite{EPJC73p2351}. Much
more straightforward is to consider $\chi_{c1}(3872)$ a unitarised and so
mass-shifted $\chi_{c1}(2P)$ state, a consequence of its naturally large
coupling to the $S$-wave $D^0\bar{D}^{\star0}$ threshold. In
Ref.~\cite{EPJC71p1762} a successful description was indeed achieved in a
momentum-space multichannel calculation with all the relevant meson-meson
channels included. The simplified coordinate-space calculation carried out
in Ref.~\cite{EPJC73p2351} aimed at studying the molecular scenario with
a two-component wave function ($c\bar{c}$ and $D^0\bar{D}^{\star0}$). 
The conclusion was that in the inner region both components are of similar
magnitude, although the $D^0\bar{D}^{\star0}$ one strongly dominates the
overall probability due to its very long tail for the then listed \cite{PDG10}
small binding energy of 0.16~MeV. Also, the $\chi_{c1}(3872)$ pole was found
both as a dynamical resonance and an intrinsic one, depending on details of
the bare $c\bar{c}$ state's mass. Finally, in Ref.~\cite{EPJC75p26} the
latter $r$-space model was generalised so as to include all relevant OZI
channels, in order to obtain realistic multicomponent wave functions of
$J/\psi$, $\psi(2S)$ and $\chi_{c1}(3872)$, allowing to compute the
measured \cite{PDG14} electromagnetic transitions
$\chi_{c1}(3872)\to J/\psi , \psi(2S)$. The resulting $\chi_{c1}(3872)$
wave function is shown in Fig.~1. We see that now the $c\bar{c}$ component
is clearly dominant in the interior region, although the $D^0\bar{D}^{\star0}$
component still accounts for most of the total probability
(65\% \cite{EPJC75p26}) because of its mentioned long tail. The
latter probability will even be close to 99\% for the recently \cite{PDG18}
adjusted $\chi_{c1}(3872)$ binding of only about 0.01~MeV. Yet the ratio of
the $c\bar{c}$ and $D^0\bar{D}^{\star0}$ components in the inner region will
hardly change \cite{EPJC73p2351}. From all this we conclude that
$\chi_{c1}(3872)$ should not be considered a meson-meson molecule
\cite{EPJC73p2351}. More generally, the effects of unitarisation are often
underestimated and should really be taken into account in modern meson
spectroscopy.
\begin{figure}[h]
\begin{center}
\includegraphics[trim = 0mm 1.5mm 0mm 0.5mm,clip,width=0.8\columnwidth]
{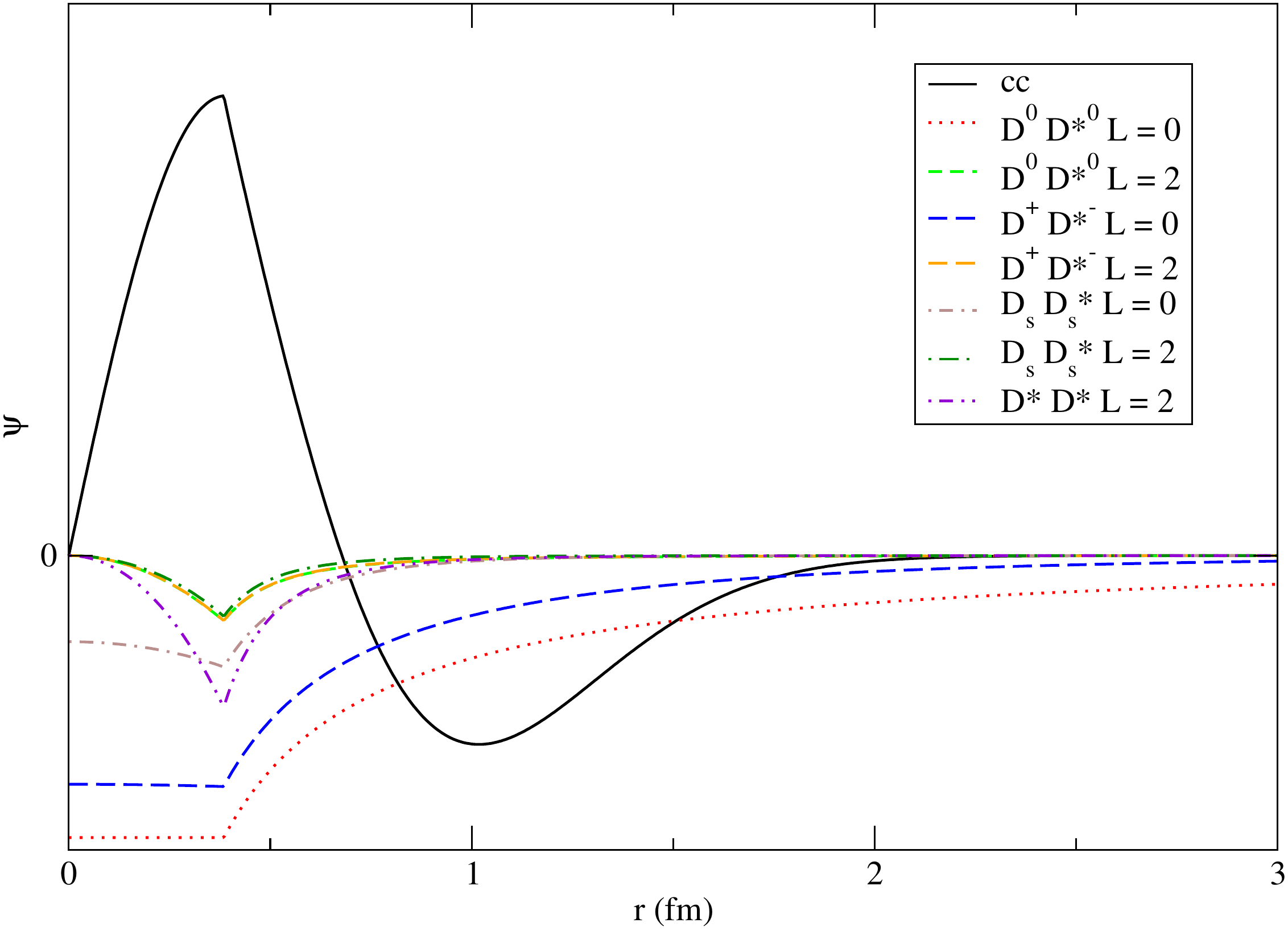}
\caption{Wave-function components of $\chi_{c1}(3872)$ in model of
Ref.~\cite{EPJC75p26}.}
\end{center}
\end{figure}
\section{Non-resonant $\bm{\psi(4260)}$, $\bm{\psi(4660)}$ and
$\bm{\Upsilon(10580)}$}
To conclude, we very succinctly review three heavy quarkonia, listed
as established $\psi$ and $\Upsilon$ resonances by the PDG \cite{PDG18},
in an unconventional way, viz.\ through data analyses that focus on threshold
effects. The latter are also usually ignored in hadron spectroscopy.
\subsection{$\bm{\psi(4260)}$}
The $\psi(4260)$ resonance, still called $X(4260)$ up to 2016 \cite{PDG16},
has 47 listed \cite{PDG18} decay modes, of which 38 are \em ``not seen''. \em
No decay into pairs of charmed mesons has ever been observed, with 
$J/\psi\pi^+\pi^-$ being the principal mode. Instead of the usual explanations
in terms of (crypto-)exotic non-$c\bar{c}$ configurations, we looked
\cite{PRL105p102001} more closely at the data. Assuming a very broad threshold
structure in the $J/\psi\pi^+\pi^-$ channel dominated by $J/\psi f_0(500)$,
we interpreted the various dips in the $J/\psi\pi^+\pi^-$ event distribution
as strong inelasticity effects from the opening of OZI-allowed channels with
pairs of charmed mesons as well as established vector charmonia in these channels.
This also helps to explain the very pronounced and puzzling dip precisely at
the mass of $\psi(4415)$. Thereabove, the opening of the
$\Lambda_c\bar{\Lambda}_c$ threshold and a tentative, so far unlisted
$\psi(3D)$ resonance are fundamental to explain the data. These large
inelasticity effects we called \em depletion, \em which give rise to the
resonance-like but --- in our opinion --- non-resonant $\psi(4260)$
enhancement.
\subsection{$\bm{\psi(4660)}$}
The situation with $\psi(4660)$ \cite{PDG18} ($X(4660)$ up until 2016
\cite{PDG16}) is related, as we consider \cite{EPL85p61002} it a non-resonant
threshold enhancement due to the opening of the $\Lambda_c\bar{\Lambda}_c$ channel
(also see Ref.~\cite{PRL105p102001}). We also found \cite{EPL85p61002}
indications of the higher vector charmonia $\psi(5S)$, $\psi(4D)$, $\psi(6S)$ and
$\psi(5D)$. As mentioned above, we identified a $\psi(3D)$ signal in 
Ref.~\cite{PRL105p102001}.
\subsection{$\bm{\Upsilon(10580)}$}
Finally, we discuss a very well established bottomonium state, namely
$\Upsilon(10580)$, included in the PDG tables since many years and listed as
$\Upsilon(4S)$. However, we interpret \cite{ARXIV10101401} it as a
non-resonant threshold bump right between the $BB$ and $BB^\star$
thresholds. The crucial point is that 
the data show a small yet clear enhancement on top of the $B_s B_s$
threshold, whereas there are unmistakable dips at the openings of the
$BB^\star$ and $B^\star B^\star$ thresholds (see Fig.~2). This pattern can
be understood by assuming an $\Upsilon(4S)$ resonance somewhat above the
$B_s B_s$ threshold, which also allows to indentify $\Upsilon(10860)$ as 
$\Upsilon(3D)$ and $\Upsilon(11020)$ as $\Upsilon(5S)$.
\begin{figure}[h]
\begin{center}
\includegraphics[trim = 0mm 1mm 0mm 1mm,clip,width=0.8\columnwidth]
{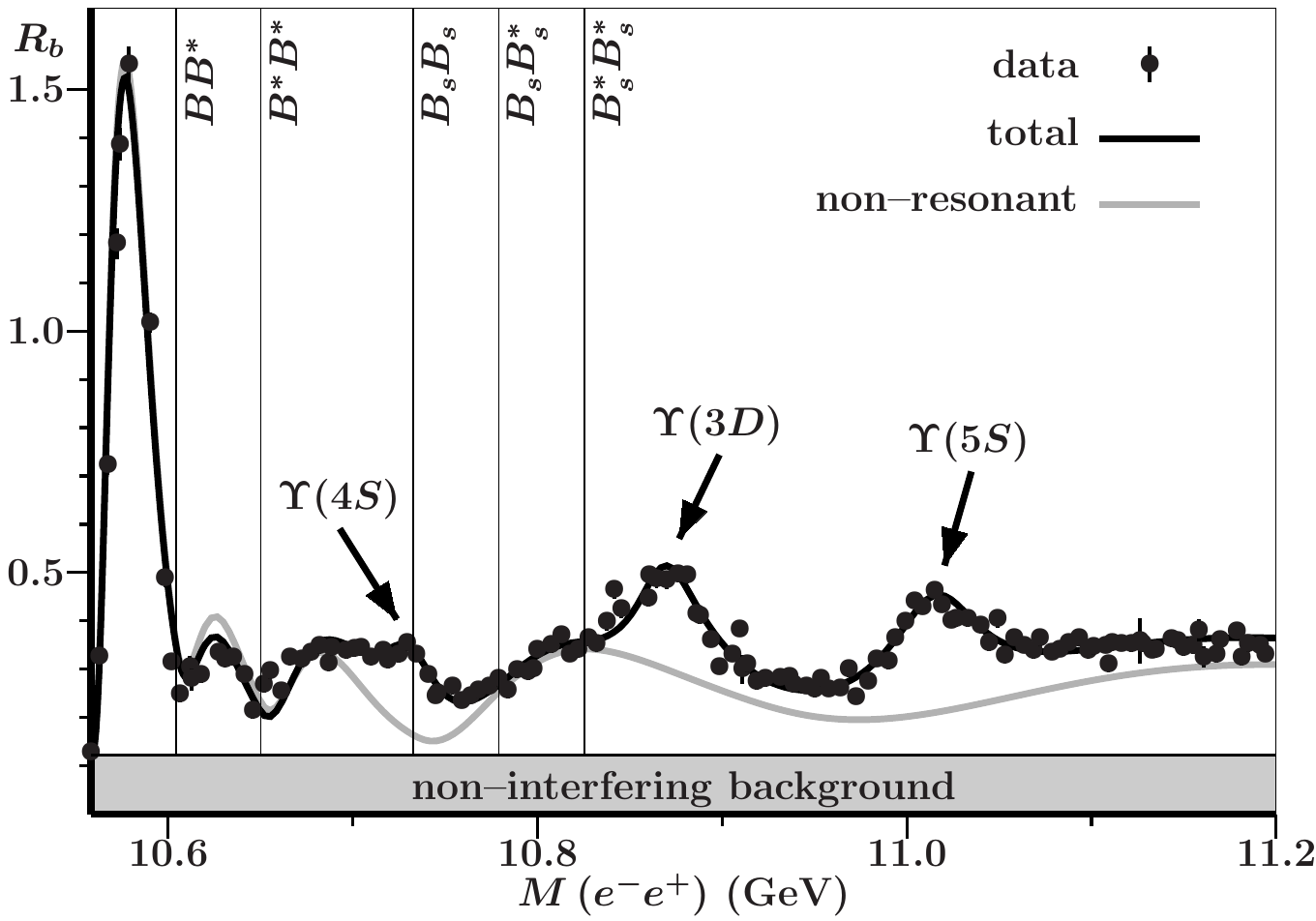}
\caption{Interpretation \cite{ARXIV10101401} of bottomonium vector states,
including non-resonant $\Upsilon(10580)$.}
\end{center}
\end{figure}

\end{document}